\documentclass[reprint,prd,aps,showpacs]{revtex4-1}

\usepackage{graphicx}

\begin{document}

\title{Nonextensive statistics based on Landsberg-Vedral entropy}

\author{A.S.~Parvan}

\affiliation{Bogoliubov Laboratory of Theoretical Physics, Joint Institute for Nuclear Research, Dubna, Russia}

\affiliation{Department of Theoretical Physics, Horia Hulubei National Institute of Physics and Nuclear Engineering, Bucharest-Magurele, Romania}

\affiliation{Institute of Applied Physics, Moldova Academy of Sciences, Chisinau, Republic of Moldova}

\begin{abstract}
The general formalism for the nonextensive statistics based on the Landsberg-Vedral entropy was derived. The formula for the first law of thermodynamics and the exact relations of the thermodynamic quantities to their ensemble averages were obtained. It was found that under the transformation $q\to 2-q$ the probabilities of microstates of the nonextensive statistics based on the Landsberg-Vedral entropy formally resemble the corresponding probabilities of the Tsallis statistics with escort probabilities. However, the nonextensive statistics with the Landsberg-Vedral entropy does not require introduction of the escort probabilities and generalized expectation values which are used in this version of the Tsallis statistics.
\end{abstract}


\maketitle

\section{Introduction}~\label{sec:1}
Initially, the equilibrium statistical mechanics was implemented on the basis of the Boltzmann entropy, the great formula $$S=k_{B}\ln W$$ that established the connection between the macroscopic variable, entropy, and the number of microstates of a system~\cite{Boltzmann1,Boltzmann2} (see also ref.~\cite{Wehrl}) and on the basis of the more general Gibbs formula $$S=-k_{B}\sum_{i} p_{i}\ln p_{i}$$ that is the entropy or "uncertainty" of the probability distribution for the microstates of a system~\cite{Jaynes1}. Lately, the Shannon expression~\cite{Shannon} for the information entropy of the discrete probability distribution allowed E.T.~Jaynes to give another formulation of the equilibrium statistical mechanics based on the maximum entropy principle or the second law of thermodynamics~\cite{Jaynes2}. The R\'{e}nyi alternative method of defining the information entropy~\cite{Renyi1,Renyi2} opened a wide way for the introduction in the information theory of new definitions of statistical entropy (see, for instance, ref.~\cite{Wehrl}). They had a significant impact on the development of the statistical models beyond the Boltzmann-Gibbs equilibrium statistical mechanics. The paper of C.~Tsallis~\cite{Tsal88} established a new direction in science showing that on the basis of the Havrda-Charv\'{a}t-Daroczy-Tsallis parametric entropy~\cite{Havrda,Daroczy,Wehrl} it is possible to construct the stationary statistical mechanics.

The Tsallis nonextensive statistics~\cite{Tsal88,Tsal98} has received a wide recognition since it is confirmed by the experiment especially in high energy physics. At the same time, the Boltzmann-Gibbs statistics faces difficulties in describing real phenomena, for example, in nuclear processes at high energies since equilibrium statistical mechanics is a simplified ideal limiting case of the behavior of reality~\cite{Jaynes2}. The equilibrium theory neglects complex processes and phenomena which take place in real conditions. However, nonextensive stationary statistics such as the Tsallis statistics covers these complex phenomena and describes them parametrically due to the fact that a theory like that goes beyond the framework of the idealized Boltzmann-Gibbs statistics. The nonextensive parametrization may indicate the presence of complex processes in real phenomena such as, for example, nontrivial macromotions, particle transformation reactions, jets, internal currents in the system and etc. Nevertheless, the Tsallis statistics may have some indeterminacy in its formulation related to the generalized expectation values and the normalization condition for the probabilities of microstates~\cite{Tsal98,Plastino1,Abe1,Parvan1}. Therefore, the main aim of this paper is to introduce the nonextensive statistics based on the Landsberg-Vedral entropy~\cite{Landsberg1} which may be formulated on the basis of the standard linear expectation values consistent with the normalization condition for the probabilities of microstates.

The Landsberg-Vedral entropy is defined as~\cite{Landsberg1}
\begin{equation}\label{1}
  S=\frac{k_{B}}{1-q} \left(1-\frac{1}{\sum\limits_{i} p_{i}^{q}}\right), \qquad q>0,
\end{equation}
where $p_{i}$ is the probability of the $i$-th microscopic state of the system. The main important properties of the Landsberg-Vedral entropy that can be mentioned are: positivity $S\geq 0$, concavity in $\{p_{i}\}$, the Gibbs limit $\lim_{q\to 1} S=-k_{B}\sum_{i} p_{i}\ln p_{i}$, and the peculiar nonextensive additivity rule~\cite{Landsberg1}
$$
S(A+B)=S(A)+S(B) + \frac{q-1}{k_{B}}S(A)S(B),
$$
for two independent events $A$ and $B$.

The structure of the paper is as follows. In sects.~\ref{sec:2}, \ref{sec:3} and \ref{sec:4}, we formulate the general theory for the nonextensive statistics based on the Landsberg-Vedral entropy in the microcanonical, canonical and grand canonical ensembles, respectively. The main conclusions are summarized in the final section.

\section{Microcanonical ensemble}\label{sec:2}
The thermodynamic potential of the microcanonical ensemble is the entropy. Let us consider the Landsberg-Vedral entropy (\ref{1}) which was introduced in~\cite{Landsberg1}. The probabilities of microstates are constrained by the additional function
\begin{equation}\label{2}
    \varphi=\sum\limits_{i}  p_{i} - 1 = 0.
\end{equation}
Then, in the microcanonical ensemble the unknown probabilities $\{p_{i}\}$ of microstates are obtained from the constrained local extrema of the thermodynamic potential (\ref{1}) by the method of the Lagrange multipliers~\cite{Krasnov}
\begin{eqnarray}\label{3}
 \Phi &=& S - \lambda \varphi,  \\ \label{4}
  \frac{\partial \Phi}{\partial p_{i}} &=& 0,
\end{eqnarray}
where $\lambda$ is an arbitrary real constant. Substituting eqs.~(\ref{1}) and (\ref{2}) into eq.~(\ref{3}) and using eqs.~(\ref{4}) and (\ref{2}), we obtain
\begin{eqnarray}\label{5}
p_{i} &=& \frac{1}{W}, \\ \label{6}
    W &=& \sum\limits_{i} \delta_{E,E_{i}}\delta_{V,V_{i}}\delta_{V,V_{i}} ,
\end{eqnarray}
where $E_{i}$, $V_{i}$ and $N_{i}$ are energy, volume and number of particles, respectively, in the $i$-th microscopic state of the system. Then, the thermodynamic potential (\ref{1}) can be rewritten as (cf. refs.~\cite{Boltzmann1,Boltzmann2,Parvan2})
\begin{equation}\label{7}
  S=\frac{W^{q-1}-1}{q-1}=z(W^{1/z}-1), \quad z\equiv\frac{1}{q-1}.
\end{equation}
Note that here and throughout the paper we use the system of natural units $\hbar=c=k_{B}=1$.

Let us suppose that $z$ is a variable of state of the system. Then, we obtain
\begin{equation}\label{8}
  dS=\frac{dz}{z} (S-W^{1/z}\ln W ) + W^{1/z} d\ln W.
\end{equation}
The statistical weight $W=W(E,V,N)$. Therefore, we have (cf. ref.~\cite{Parvan2} for the Tsallis statistics)
\begin{equation}\label{9}
  TdS=dE+pdV+Xdz-\mu dN,
\end{equation}
where
\begin{eqnarray}\label{10}
  \frac{1}{T} &=& W^{1/z} \frac{\partial \ln W}{\partial E} = \left(\frac{\partial S}{\partial E}\right)_{VzN},  \\ \label{11}
  \frac{p}{T} &=& W^{1/z} \frac{\partial \ln W}{\partial V} = \left(\frac{\partial S}{\partial V}\right)_{EzN}, \\ \label{12}
  -\frac{\mu}{T} &=& W^{1/z} \frac{\partial \ln W}{\partial N} = \left(\frac{\partial S}{\partial N}\right)_{EVz}
\end{eqnarray}
and
\begin{equation}\label{13}
  \frac{X}{T} = W^{1/z}-1-W^{1/z}\ln W^{1/z}= \left(\frac{\partial S}{\partial z}\right)_{EVN}.
\end{equation}
The first partial derivatives of the thermodynamic potential $S$ with respect to the variables of state in eqs.~(\ref{10})--(\ref{13}) can also be verified from eq.~(\ref{7}). Equation (\ref{9}) represents the first law of thermodynamics. Note that in the Gibbs limit $q\to 1$ we obtain all the relations of the Boltzmann-Gibbs statistics in the microcanonical ensemble.

\section{Canonical ensemble}\label{sec:3}
The thermodynamic potential of the canonical ensemble, the free energy $F$, is obtained from the fundamental thermodynamic potential, the energy $E$, by the Legendre transform. Using eq.~(\ref{1}), we have
\begin{equation}\label{14}
  F=E-TS=\sum\limits_{i} p_{i} E_{i}-\frac{T}{1-q} \left(1-\frac{1}{\sum\limits_{i} p_{i}^{q}}\right),
\end{equation}
where $E=\sum\limits_{i} p_{i} E_{i}$. In the canonical ensemble the Lagrange function can be written as~\cite{Parvan2015a}:
\begin{equation}\label{15}
  \Phi = F - \lambda \varphi,
\end{equation}
where $\lambda$ is an arbitrary real constant. Substituting eqs.~(\ref{14}) and (\ref{2}) into eq.~(\ref{15}) and using eqs.~(\ref{4}) and (\ref{2}), we obtain
\begin{eqnarray}\label{16}
p_{i} &=& \frac{1}{Z_{c}}\left[1+(q-1)\frac{\Lambda-E_{i}}{T}\right]^{\frac{1}{q-1}}, \\ \label{17}
    Z_{c} &=& \sum\limits_{i}  \left[1+(q-1)\frac{\Lambda-E_{i}}{T}\right]^{\frac{1}{q-1}}
\end{eqnarray}
and
\begin{equation}\label{18}
  Z_{c}^{q-1}=\frac{q}{\chi^{2}}, \quad \chi\equiv \sum\limits_{i} p_{i}^{q},
\end{equation}
where $\Lambda=\lambda-T/(q-1)$. Substituting eq.~(\ref{16}) into $\chi$ and using eqs.~(\ref{17}), (\ref{18}), we find an equation for $\Lambda$ as
\begin{eqnarray}\label{19}
&& \sum\limits_{i}  \left[1+(q-1)\frac{\Lambda-E_{i}}{T}\right]^{\frac{1}{q-1}}    \nonumber \\
 &&  = \left\{q^{-1/2}\sum\limits_{i} \left[1+(q-1)\frac{\Lambda-E_{i}}{T}\right]^{\frac{q}{q-1}} \right\}^{\frac{2}{q+1}} .
\end{eqnarray}

Substituting eq.~(\ref{16}) into eq.~(\ref{1}) and using eq.~(\ref{2}), we have
\begin{equation}\label{20}
  S=\frac{1}{1+(q-1)\frac{\Lambda-E}{T}}\left[\frac{Z_{c}^{q-1}-1}{q-1}-\frac{\Lambda-E}{T}\right].
\end{equation}
Then, the thermodynamic potential (\ref{14}) can be written as
\begin{eqnarray}\label{21}
  F &=& \Lambda-\frac{T}{1+(q-1)\frac{\Lambda-E}{T}} \nonumber \\
  &\times& \left[\frac{Z_{c}^{q-1}-1}{q-1}+(q-1)\left(\frac{\Lambda-E}{T}\right)^{2}\right].
\end{eqnarray}

In the canonical ensemble the entropy $S=S(T,V,z,N)$. Thus, using eq.~(\ref{1}) and the definition of $z$ given in eq.~(\ref{7}), we obtain
\begin{equation}\label{22}
  dS= \frac{dz}{z} \left(S+\frac{1}{\chi^{2}}\sum\limits_{i} p_{i}^{1+\frac{1}{z}}\ln p_{i} \right) - \frac{z (1+\frac{1}{z})}{\chi^{2}}
  \sum\limits_{i} p_{i}^{\frac{1}{z}} dp_{i}.
\end{equation}
Substituting eq.~(\ref{16}) into eq.~(\ref{22}) and using $\sum\limits_{i} dp_{i}=0$ and eq.~(\ref{18}), we can write
\begin{equation}\label{23}
  dS= \frac{dz}{z} \left(S+\frac{1}{\chi^{2}}\sum\limits_{i} p_{i}^{1+\frac{1}{z}}\ln p_{i} \right) + \frac{1}{T} \sum\limits_{i} E_{i} dp_{i}.
\end{equation}
The differential of the energy $E$ can be written as
\begin{eqnarray}\label{24}
dE &=& \sum\limits_{i} p_{i} dE_{i} + \sum\limits_{i} E_{i} dp_{i}, \\ \label{25}
    dE_{i} &=& \frac{\partial E_{i}}{\partial V} dV + \frac{\partial E_{i}}{\partial N} dN,
\end{eqnarray}
where $\partial E_{i}/\partial T =\partial E_{i}/\partial z =0$. Then, substituting eqs.~(\ref{24}) and (\ref{25}) into eq.~(\ref{23}), we obtain the first law of thermodynamics (cf. ref.~\cite{Parvan3} for the Tsallis statistics)
\begin{equation}\label{26}
  TdS=dE+pdV+Xdz-\mu dN,
\end{equation}
where
\begin{eqnarray}\label{27}
  p &=& -\sum\limits_{i} p_{i} \frac{\partial E_{i}}{\partial V}, \\ \label{28}
 \mu &=& \sum\limits_{i} p_{i} \frac{\partial E_{i}}{\partial N}
\end{eqnarray}
and
\begin{equation}\label{29}
  X =\frac{T}{z} \left(S+\frac{1}{\chi^{2}}\sum\limits_{i} p_{i}^{1+\frac{1}{z}}\ln p_{i} \right).
\end{equation}
Then, from eq.~(\ref{26}) and the Legendre transform (\ref{14}) the differential thermodynamic relation for the thermodynamic potential $F$ can be written as
\begin{equation}\label{30}
  dF=-SdT-pdV-Xdz+\mu dN.
\end{equation}

Taking the first derivative of the thermodynamic potential (\ref{14}) with respect to temperature $T$ and using eqs.~(\ref{16}) and (\ref{18}), and the relations $\partial E_{i}/\partial T=0$, $\sum_{i} \partial p_{i}/\partial T=0$, we obtain the entropy $S$ of the system as (cf. eq.~(\ref{1}))
\begin{equation}\label{31}
   S =-\left(\frac{\partial F}{\partial T}\right)_{VzN} = \frac{1}{1-q} \left(1-\frac{1}{\sum\limits_{i} p_{i}^{q}}\right).
\end{equation}
Taking the first derivative of the thermodynamic potential (\ref{14}) with respect to volume $V$ and using eqs.~(\ref{16}) and (\ref{18}), and the relation $\sum_{i} \partial p_{i}/\partial V=0$, we obtain the pressure $p$ of the system as (cf. eq.~(\ref{27}))
\begin{equation}\label{32}
   p =-\left(\frac{\partial F}{\partial V}\right)_{TzN} = -\sum\limits_{i} p_{i} \frac{\partial E_{i}}{\partial V}.
\end{equation}
Taking the first derivative of the thermodynamic potential (\ref{14}) with respect to the number of particles $N$ and using eqs.~(\ref{16}) and (\ref{18}), and the relation $\sum_{i} \partial p_{i}/\partial N=0$, we obtain the chemical potential $\mu$ of the system as (cf. eq.~(\ref{28}))
\begin{equation}\label{33}
   \mu =\left(\frac{\partial F}{\partial N}\right)_{TVz} = \sum\limits_{i} p_{i} \frac{\partial E_{i}}{\partial N}.
\end{equation}
Taking the first derivative of the thermodynamic potential (\ref{14}) with respect to variable $z$ and using eqs.~(\ref{16}) and (\ref{18}), and the relations $\partial E_{i}/\partial z=0$, $\sum_{i} \partial p_{i}/\partial z=0$, we obtain the conjugate force $X$ of the system as (cf. eq.~(\ref{29}))
\begin{equation}\label{34}
   X =-\left(\frac{\partial F}{\partial z}\right)_{TVN} = \frac{T}{z} \left(S+\frac{1}{\chi^{2}}\sum\limits_{i} p_{i}^{1+\frac{1}{z}}\ln p_{i} \right).
\end{equation}
Applying the Legendre back-transformation to the function $F$ and using eqs.~(\ref{14}) and (\ref{31}), we obtain the mean energy of the system as
\begin{equation}\label{35}
  E = -T^{2} \left(\frac{\partial}{\partial T} \frac{F}{T}\right)_{VzN} = \sum\limits_{i}  p_{i} E_{i}.
\end{equation}
Note that the ensemble averages for the thermodynamic quantities (\ref{32}), (\ref{33}) and (\ref{35}) in the canonical ensemble are the same as those of the Boltzmann-Gibbs statistics with the exception of the probability of microstates which, here, is given by eq.~(\ref{16}).

Let us rewrite the probabilities of microstates (\ref{16}) in another representation. Substituting eq.~(\ref{16}) into the function $\chi$ defined in eq.~(\ref{18}) and using the relation for $Z_{c}$ given also in eq.~(\ref{18}), we obtain
\begin{equation}\label{36}
  1+(q-1)\frac{\Lambda}{T}=\frac{q}{\chi}+(q-1)\frac{E}{T}.
\end{equation}
Substituting eq.~(\ref{36}) into eqs.~(\ref{16}), (\ref{17}), we have
\begin{eqnarray}\label{37}
p_{i} &=& \frac{1}{Z_{q}}\left[1+(q-1)\frac{E-E_{i}}{cT}\right]^{\frac{1}{q-1}}, \\ \label{38}
    Z_{q} &=& \sum\limits_{i} \left[1+(q-1)\frac{E-E_{i}}{cT}\right]^{\frac{1}{q-1}}
\end{eqnarray}
and
\begin{equation}\label{39}
 c=q Z_{q}^{q-1},
\end{equation}
where $\chi=q/c$. It can be observed that under the transformation $q\to 2-q$ the probability of microstates (\ref{37}) formally recovers the probability of microstates of the Tsallis statistics with escort probabilities~\cite{Tsal98,Abe1999}. However, there is a cardinal difference between the present statistics and the Tsallis statistics with escort probabilities. In the present statistics based on the Landsberg-Vedral entropy we use the physical standard expectation values, i.e. the mean values of the dynamical variables that have a linear dependence on the probabilities of microstates which are consistent with the normalization condition for the probabilities of microstates. However, the Tsallis statistics with escort probabilities deals with the generalized expectation values which are defined on the base of the escort probabilities. Note that in the Gibbs limit $q\to 1$ we obtain all the relations of the Boltzmann-Gibbs statistics in the canonical ensemble.

\section{Grand canonical ensemble}\label{sec:4}
The thermodynamic potential of the grand canonical ensemble is related to the fundamental thermodynamic potential, energy $E$, by the Legendre transform. Using eq.~(\ref{1}), we have
\begin{equation}\label{40}
 \Omega = E - TS - \mu N =\sum\limits_{i} p_{i} (E_{i}-\mu N_{i})-\frac{T}{1-q} \left(1-\frac{1}{\sum\limits_{i} p_{i}^{q}}\right),
\end{equation}
where $E=\sum_{i} p_{i} E_{i}$ and $N=\sum_{i} p_{i} N_{i}$. In the grand canonical ensemble the Lagrange function can be written as~\cite{Parvan1}:
\begin{equation}\label{41}
  \Phi = \Omega - \lambda \varphi,
\end{equation}
where $\lambda$ is an arbitrary real constant. Substituting eqs.~(\ref{40}) and (\ref{2}) into eq.~(\ref{41}) and using eqs.~(\ref{4}) and (\ref{2}), we obtain
\begin{eqnarray}\label{42}
p_{i} &=& \frac{1}{Z_{\Omega}}\left[1+(q-1)\frac{\Lambda-E_{i}+\mu N_{i}}{T}\right]^{\frac{1}{q-1}}, \\ \label{43}
    Z_{\Omega} &=& \sum\limits_{i}  \left[1+(q-1)\frac{\Lambda-E_{i}+\mu N_{i}}{T}\right]^{\frac{1}{q-1}}
\end{eqnarray}
and
\begin{equation}\label{44}
  Z_{\Omega}^{q-1}=\frac{q}{\chi^{2}},
\end{equation}
where $\chi$ and $\Lambda$ are defined in eq.~(\ref{18}) and below it, respectively. Substituting eq.~(\ref{42}) into $\chi$ and using eqs.~(\ref{43}) and (\ref{44}), we found an equation for $\Lambda$ as
\begin{eqnarray}\label{45}
&& \sum\limits_{i}  \left[1+(q-1)\frac{\Lambda-E_{i}+\mu N_{i}}{T}\right]^{\frac{1}{q-1}}    \nonumber \\
 &&  = \left\{q^{-1/2}\sum\limits_{i} \left[1+(q-1)\frac{\Lambda-E_{i}+\mu N_{i}}{T}\right]^{\frac{q}{q-1}} \right\}^{\frac{2}{q+1}}. \;\;\;\;\;
\end{eqnarray}

Substituting eq.~(\ref{42}) into eq.~(\ref{1}) and using eq.~(\ref{2}), we have
\begin{equation}\label{46}
  S=\frac{1}{1+(q-1)\frac{\Lambda-E+\mu N}{T}}\left[\frac{Z_{\Omega}^{q-1}-1}{q-1}-\frac{\Lambda-E+\mu N}{T}\right].
\end{equation}
Then, the thermodynamic potential (\ref{40}) can be rewritten as
\begin{eqnarray}\label{47}
  \Omega &=& \Lambda-\frac{T}{1+(q-1)\frac{\Lambda-E+\mu N}{T}} \nonumber \\
  &\times& \left[\frac{Z_{\Omega}^{q-1}-1}{q-1}+(q-1)\left(\frac{\Lambda-E+\mu N}{T}\right)^{2}\right].
\end{eqnarray}

In the grand canonical ensemble the entropy $S=S(T,V,z,\mu)$. Thus, using eq.~(\ref{1}) and the definition of $z$ given in eq.~(\ref{7}), we obtain exactly eq.~(\ref{22}).
Substituting eq.~(\ref{42}) into eq.~(\ref{22}) and using $\sum_{i} dp_{i}=0$ and eq.~(\ref{44}), we obtain
\begin{equation}\label{48}
  dS= \frac{X}{T} dz + \frac{1}{T} \sum\limits_{i} (E_{i}-\mu N_{i}) dp_{i},
\end{equation}
where $X$ is the same as in eq.~(\ref{29}). The differential of the mean energy $E$ and the mean number of particles $N$ can be written as
\begin{eqnarray}\label{49}
dE-\mu dN &=& \sum\limits_{i} p_{i} (dE_{i}-\mu dN_{i}) \nonumber \\
 &+& \sum\limits_{i} (E_{i}-\mu N_{i}) dp_{i}, \\ \label{50}
  dE_{i}-\mu dN_{i} &=& \left(\frac{\partial E_{i}}{\partial V} -\mu \frac{\partial N_{i}}{\partial V}\right) dV,
\end{eqnarray}
where $\partial E_{i}/\partial T = \partial E_{i}/\partial z =\partial E_{i}/\partial \mu =0$ and $\partial N_{i}/\partial T = \partial N_{i}/\partial z =\partial N_{i}/\partial \mu =0$. Then, substituting eqs.~(\ref{49}) and (\ref{50}) into eq.~(\ref{48}), we obtain the first law of thermodynamics
\begin{equation}\label{51}
  TdS=dE+pdV+Xdz-\mu dN,
\end{equation}
where
\begin{equation}\label{52}
  p = -\sum\limits_{i} p_{i} \left(\frac{\partial E_{i}}{\partial V} -\mu \frac{\partial N_{i}}{\partial V}\right).
\end{equation}
Then, from eq.~(\ref{51}) and the Legendre transform (\ref{40}) the differential thermodynamic relation for the thermodynamic potential $\Omega$ can be written as (cf. ref.~\cite{Parvan1} for the Tsallis statistics)
\begin{equation}\label{53}
  d\Omega=-SdT-pdV-Xdz-Nd\mu.
\end{equation}

Taking the first derivative of the thermodynamic potential (\ref{40}) with respect to temperature $T$ and using eqs.~(\ref{42}) and (\ref{44}), and the relations $\partial E_{i}/\partial T=\partial N_{i}/\partial T =0$, $\sum_{i} \partial p_{i}/\partial T=0$, we obtain the entropy $S$ of the system as (cf. eq.~(\ref{1}))
\begin{equation}\label{54}
   S =-\left(\frac{\partial \Omega}{\partial T}\right)_{Vz\mu} = \frac{1}{1-q} \left(1-\frac{1}{\sum\limits_{i} p_{i}^{q}}\right).
\end{equation}
Taking the first derivative of the thermodynamic potential (\ref{40}) with respect to volume $V$ and using eqs.~(\ref{42}) and (\ref{44}), and the relation $\sum_{i} \partial p_{i}/\partial V=0$, we obtain the pressure $p$ of the system as (cf. eq.~(\ref{52}))
\begin{equation}\label{55}
   p =-\left(\frac{\partial \Omega}{\partial V}\right)_{Tz\mu} = -\sum\limits_{i} p_{i} \left(\frac{\partial E_{i}}{\partial V} -\mu \frac{\partial N_{i}}{\partial V}\right).
\end{equation}
Taking the first derivative of the thermodynamic potential (\ref{40}) with respect to the chemical potential $\mu$ and using eqs.~(\ref{42}) and (\ref{44}), and the relations $\partial E_{i}/\partial \mu=\partial N_{i}/\partial \mu =0$, $\sum_{i} \partial p_{i}/\partial \mu=0$, we obtain the mean number of particles $N$ of the system as
\begin{equation}\label{56}
   N =-\left(\frac{\partial \Omega}{\partial \mu}\right)_{TVz} = \sum\limits_{i} p_{i} N_{i}.
\end{equation}
Taking the first derivative of the thermodynamic potential (\ref{40}) with respect to variable $z$ and using eqs.~(\ref{42}) and (\ref{44}), and the relations $\partial E_{i}/\partial z=\partial N_{i}/\partial z=0$, $\sum_{i} \partial p_{i}/\partial z=0$, we obtain the conjugate force $X$ of the system as (cf. eq.~(\ref{29}))
\begin{equation}\label{57}
   X =-\left(\frac{\partial \Omega}{\partial z}\right)_{TV\mu} = \frac{T}{z} \left(S+\frac{1}{\chi^{2}}\sum\limits_{i} p_{i}^{1+\frac{1}{z}}\ln p_{i} \right).
\end{equation}
Applying the Legendre back-transformation to the function $\Omega$ and using eqs.~(\ref{40}) and (\ref{54}), we obtain the mean energy of the system as
\begin{equation}\label{58}
  E = -T^{2} \left(\frac{\partial}{\partial T} \frac{\Omega}{T}\right)_{Vz\mu} +\mu N = \sum\limits_{i}  p_{i} E_{i}.
\end{equation}
It should be stressed that the ensemble averages for the thermodynamic quantities (\ref{55}), (\ref{56}) and (\ref{58}) in the grand canonical ensemble are the same as those of the Boltzmann-Gibbs statistics with the exception of the probability of microstates which, here, is given by eq.~(\ref{42}). In the Gibbs limit $q\to 1$ we obtain all the relations of the Boltzmann-Gibbs statistics in the grand canonical ensemble.

\section{Conclusions}\label{sec:5}
In conclusion, we have introduced new nonextensive statistics based on the Landsberg-Vedral entropy. In its formulation we have used the standard linear expectation values constrained with the standard normalization condition for the probabilities of microscopic states of the system. The three ensembles were considered: microcanonical, canonical and grand canonical. The probabilities of microstates were obtained from the principle of maximum entropy using the method of the Lagrange multipliers. In each of these three ensembles we have derived exactly the first law of thermodynamics. We have also obtained the exact relations between the thermodynamic definitions of the thermodynamic quantities and their ensemble averages. The ensemble averages for the pressure, the chemical potential, the mean number of particles and the mean energy in both canonical and grand canonical ensembles are the same as those of the Boltzmann-Gibbs statistics with the exception of the form of the probability of microstates. In this formalism the Legendre transform is preserved. We have found that under the transformation $q\to 2-q$ the probabilities of microstates for the nonextensive statistics based on the Landsberg-Vedral entropy formally recover the probabilities of microstates of the Tsallis statistics with escort probabilities. However, the nonextensive statistics based on the Landsberg-Vedral entropy does not require introduction of the complicated escort probabilities and generalized expectation values which lie in the definition of the Tsallis statistics with escort probabilities.

{\bf Acknowledgments:} This work was supported in part by the joint research project of JINR and IFIN-HH (protocol N 4543).


\begin{thebibliography}{}
%
\bibitem{Boltzmann1} L.~Boltzmann, Wiener Ber. 75 (1877) 67.

\bibitem{Boltzmann2} L.~Boltzmann, Wiener Ber. 76 (1877) 373.

\bibitem{Wehrl} A.~Wehrl, Rev. Mod. Phys. 50 (1978) 221.

\bibitem{Jaynes1} E.T.~Jaynes, Am. J. Phys. 33 (1965) 391.

\bibitem{Shannon} C.E.~Shannon, Bell System Tech. J. 27 (1948) 379.

\bibitem{Jaynes2} E.T.~Jaynes, Phys. Rev. 106 (1957) 620.

\bibitem{Renyi1} A.~R\'{e}nyi, Proc. Fourth Berkeley Symposium on Mathematical Statistics and Probability Vol. I (1961) 547.

\bibitem{Renyi2} A.~R\'{e}nyi, \textit{Probability Theory} (North-Holland, Amsterdam, 1970).


\bibitem{Tsal88} C.~Tsallis, J. Stat. Phys. 52 (1988) 479.

\bibitem{Havrda} J.~Havrda, F.~Charv\'{a}t, Kybernetica 3 (1967) 30.

\bibitem{Daroczy} Z.~Daroczy, Inf. Control 16 (1970) 36.

\bibitem{Tsal98} C.~Tsallis, R.S.~Mendes, A.R.~Plastino, Physica A 261, (1998) 534.

\bibitem{Plastino1} A. Plastino, A.R. Plastino, Phys. Lett. A 226 (1997) 257.

\bibitem{Abe1} S.~Abe, Phys. Lett. A 275 (2000) 250.

\bibitem{Parvan1} A.S.~Parvan, Eur. Phys. J. A 51 (2015) 108.


\bibitem{Landsberg1} P.T.~Landsberg, V.~Vedral, Phys. Lett. A 247 (1998) 211.


\bibitem{Krasnov} M.L.~Krasnov, G.I.~Makarenko, A.I.~Kiseliov, \textit{Calculus of Variations: Problems and Exercises with detailed solutions} (URSS Publisher, Moscow 2002).

\bibitem{Parvan2} A.S.~Parvan, Phys. Lett. A 350 (2006) 331.

\bibitem{Parvan2015a} A.S.~Parvan, Foundation of equilibrium statistical mechanics based on generalized entropy, in {\it Recent Advances in Thermo and Fluid Dynamics}, ed. by Mofid Gorji-Bandpy (InTech, Rijeka, 2015), p.303.

\bibitem{Parvan3} A.S.~Parvan, Phys. Lett. A 360 (2006) 26.

\bibitem{Abe1999} S.~Abe, Physica A 269 (1999) 403.

\end{thebibliography}
\end{document}